\documentclass[preprint,showpacs,prd]{revtex4}
\usepackage{epsfig}

\begin{document}


\title{\boldmath Partial Wave Analyses of $J/\psi\to\gamma K^+K^-$ and $\gamma
K^0_SK^0_S$}

\author{\small
J.~Z.~Bai$^1$,        Y.~Ban$^{10}$,         J.~G.~Bian$^1$,
D.~V.~Bugg$^{11}$,
X.~Cai$^{1}$,          J.~F.~Chang$^1$,
H.~F.~Chen$^{17}$,    H.~S.~Chen$^1$,   
H.~X.~Chen$^{3}$,
Jie~Chen$^{9}$
% ...
\\
(BES Collaboration)
}
\vspace{0.1cm}
\affiliation{
$^1$ Institute of High Energy Physics, Beijing 100039, People's Republic of
     China\\
% ...
$^3$ Guangxi Normal University, Guilin 541004, People's Republic of China\\
% ...
$^{9}$ Nankai University, Tianjin 300071, People's Republic of China\\
$^{10}$ Peking University, Beijing 100871, People's Republic of China\\
$^{11}$ Queen Mary, London E14NS, UK \\
% ...
$^{17}$ University of Science and Technology of China, Hefei 230026,
        People's Republic of China\\
% ...
}

\date{July 21, 2003}

\begin{abstract}
  Results are presented on $J/\psi $ radiative decays to $K^+K^-$ and
  $K^0_SK^0_S$ based on a sample of 58M $J/\psi$ events taken with the BES\,II
  detector. A partial wave analysis is carried out using the
  relativistic covariant tensor amplitude method in the 1-2 GeV mass
  range.  There is conspicuous production due to the $f'_2(1525)$ and
  $f_0(1710)$.  The latter peaks at a mass of $1740\pm 4^{+10}_{-25}$
  MeV with a width of $166{^{+5}_{-8}}{^{+15}_{-10}}$ MeV.  Spin 0 is
  strongly preferred over spin 2. For the $f'_2(1525)$, the helicity
  amplitude ratios are determined to be $x^2 = 1.00\pm0.28^{+1.06}_{-0.36}$
  and $y^2 = 0.44\pm{0.08}^{+0.10}_{-0.56}$.
\end{abstract}



\maketitle


\section{Introduction}

QCD predicts the existence of glueballs, the bound states of gluons,
and the observation of glueballs is, to some extent, a direct test of
QCD. Such gluonic states are expected to give rise to a rich isoscalar
meson spectroscopy, and Lattice Gauge Theory calculations predict, in
particular, that the lowest-lying state should occur in the mass range
1.4-1.8 GeV and have $J^{PC} = 0^{++}$~\cite{QCDL}.  For a $J/\psi$
radiative decay to two pseudoscalar mesons, only $J^{PC}$ values in
the series $0^{++},~2^{++},~...$ are possible, so such states provide
a very clean laboratory to search for the lowest mass scalar glueball.

There has been a long history of uncertainty about the properties of
the $f_0(1710)$, one of the earliest glueball candidates.  This
history is reviewed in detail in the latest issue of the Particle Data
Group (PDG) \cite{PDG} and will not be repeated here.  The latest analysis
of Mark III data by Dunwoodie \cite{WMD} favors $J^P = 0^+$ over an
earlier assignment of $2^+$, while the latest central production data
of WA76 and WA102 also favor $0^+$ \cite{WA76, WA102}.  In this paper,
we present new results on $J/\psi \to \gamma K^+K^-$ and $\gamma K^0_SK^0_S$
based on a sample of 58M $J/\psi $ events taken with the upgraded
Beijing Spectrometer (BES\,II) located at the Beijing Electron
Positron Collider (BEPC).

\section{Bes detector}
BES\,II is a large
solid-angle magnetic spectrometer that is described in detail in Ref.
\cite{BESII}. Charged particle momenta are determined with a
resolution of $\sigma_p/p = 1.78\%\sqrt{1+p^2(\mbox{GeV}^2)}$ in a
40-layer cylindrical drift chamber. Particle identification is
accomplished by specific ionization ($dE/dx$) measurements in the
drift chamber and time-of-flight (TOF) measurements in a barrel-like
array of 48 scintillation counters. The $dE/dx$ resolution is
$\sigma_{dE/dx} = 8.0\%$; the TOF resolution is $\sigma_{TOF} = 180$
ps for Bhabha events. Outside of the time-of-flight counters is a
12-radiation-length barrel shower counter (BSC) comprised of gas
proportional tubes interleaved with lead sheets. The BSC measures the
energies and directions of photons with resolutions of
$\sigma_E/E\simeq 21\%/\sqrt{E(\mbox{GeV})}$, $\sigma_{\phi} = 7.9$
mrad, and $\sigma_{z}$ = 2.3 cm. The iron flux return of the
magnet is instrumented with three double layers of counters that are
used to identify muons.  The average luminosity of the BEPC
accelerator is $4.0 \times 10^{30}$ cm$^{-2}$$s^{-1}$ at the
center-of-mass energy of 3.1 GeV.

In this analysis, a
 GEANT3 based Monte Carlo simulation package (SIMBES) with detailed 
   consideration of real detector performance (such as dead
   electronic channels) is used.
    The consistency between data and Monte Carlo has been carefully checked in
   many high purity physics channels, and the agreement is quite reasonable.

\section{Event selection}
The first level of event selection requires two charged tracks with
total charge zero for $\gamma K^+K^-$ candidate events, and requires
two positively-charged and two negatively-charged tracks for
$\gamma K^0_SK^0_S$ events. These tracks are required to lie well
within the acceptance of the detector and to have a good helix fit.
More than one photon per event is allowed because of the possibility
of fake photons coming from the interactions of charged tracks
with the shower counter or from electronic noise in the shower
counter.

For $J/\psi\to\gamma K^+K^-$, the vertex is required to lie within 2
cm of the beam axis ($x-y$ plane) and within 20 cm of the center of
the interaction region (along $z$). Each of the charged particles is
required to not register hits in the muon counters in order to remove
$\gamma\mu^+\mu^-$ events. The following selection criteria are used
to remove the large backgrounds from Bhabha events: (i) The opening
angle of the two tracks satisfies $\theta_{op} < 175^{\circ}$. (ii)
The energy deposit of each track in BSC satisfies $E_{SC} < 1.0$ GeV.
In order to reduce the background from final states with pions and
electrons, each event is required to have at least one kaon identified
by the TOF.  Requirements on two variables, $U$ and $P^2_{t\gamma}$,
are imposed \cite{THimel}. A ``missing-neutral-energy'' variable $ U =
(E_{miss} - \vert \stackrel {\rightarrow} P_{miss}\vert) $ is required
to satisfy $ -0.10 < U < 0.20$ GeV; here $E_{miss}$ and $\stackrel
{\rightarrow}P_{miss} $ are the missing energy and momentum of all
charged particles respectively.  Also a ``missing-$p_t$'' variable
$P^2_{t\gamma}$ = $4\vert \stackrel {\rightarrow}P_{miss}\vert^2
\sin^2{{\theta_{\gamma}}/2}$ is required to be $< 0.002$ GeV$^2$,
where $\theta_{\gamma}$ is the angle between the missing momentum and
the photon direction.  The $U$ cut removes most background from events
having multipion or other neutral particles, such as
$\rho\pi,~\gamma\pi^+\pi^-$ events; $P^2_{t\gamma}$ is used to
eliminate background photons.  The selection criteria for a good
photon used here are based on those applied in previous BES\,I
analyses \cite{JINS}. In brief, the good photon is required to be
isolated from the two charged tracks and to come from the interaction
point.

In order to reduce the $J/\psi\to\pi^0 K^+K^-$ and
$J/\psi\to\pi^0\pi^+\pi^-$ contamination, all events surviving the
above criteria which have two or more photons are kinematically fitted to these
hypotheses. Those events with a fit $\chi^2 < 50$, and with photon
pair invariant
mass within 50 MeV$/c^2$ of the $\pi^0$ mass, are
rejected. Finally, the two charged tracks and photon in the
event are 4-C kinematically fitted to obtain better mass resolution
and to suppress backgrounds further by the requirements $\chi
^2_{\gamma K^+K^-} < 10$ and $\chi ^2_{\gamma K^+K^-} < \chi^2_{\gamma
\pi ^+\pi ^-}$.

For $J/\psi\to\gamma K^0_SK^0_S$, the $K^0_S$
mesons in the event are identified through the decay
$K^0_S\to\pi^+\pi^-$. The four charged tracks can be grouped into two
pairs, each having two oppositely charged tracks with an acceptable
distance of closest approach.
Signal events are
required to satisfy $\delta^2_{K_S} <  (20\mbox{MeV}/c^2)^2$, where $\delta^2_{K_S} =
(M_{\pi^+\pi^-}(1)-M_{K_S})^2+(M_{\pi^+\pi^-}(2)-M_{K_S})^2$ and $M_{\pi^+\pi^-}$ is calculated at the
$K^0_S$ decay vertex. The main backgrounds from $\gamma
K^0_SK^{\pm}\pi^{\mp}$ and $\gamma K^0_S K^0_S\pi^0$ events are
suppressed by requiring $U < 0.10$ GeV, $P^2_{t\gamma} < 0.005$
GeV$^2$ and the 4-C kinematic
 fit $\chi^2_{\gamma 4\pi} < 10$.

\begin{figure}
 \centerline{\epsfig{file=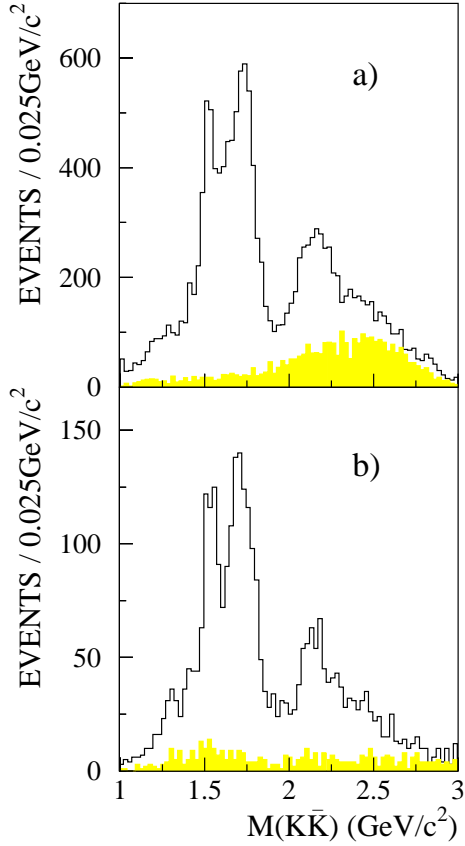,width=60mm}}
\caption{\label{fig:epsart} Invariant mass spectra of a) $K^+K^-$, b)
$K^0_SK^0_S$ for $J/\psi \to \gamma K \bar K$ events, where the shaded histograms
correspond to the estimated background contributions.}
  \label{gkkb}
\end{figure}

Fig. \ref{gkkb} shows the $K^+K^-$ and $K^0_SK^0_S$ mass spectra for
the selected events, together with the corresponding background distributions. These two mass
spectra agree closely below 2.0 GeV. The resonant structures in the
mass regions of the $f'_2(1525)$ and the $f_0(1710)$ are very clearly
visible in both decay modes.
Averaged over the whole
mass range, the detection efficiency for $\gamma K^+K^-$ is
$14.7\%$ and for $\gamma K^0_SK^0_S$ is $14.5\%$. For the $\gamma
K^+K^-$ channel,
the experimental background arises
mainly from the non-resonant $K^+K^-\pi ^0$ and
two-body $K^{*\pm}K^{\mp}$ events which are peaked at high $K^+K^-$
masses.
In the entire mass range,
14597 $\gamma K^+K^-$ events are reconstructed, and the detailed
Monte Carlo simulation
of the BES detector estimates a background of 3094 events. 
The
estimation of the background events in the $\gamma K^0_SK^0_S$ sample
is obtained from the $\delta^2_{K_S}$ side band
$(28.7\mbox{MeV}/c^2)^2 < \delta^2_{K_S} < (35\mbox{MeV}/c^2)^2$; this
equal-area-selection provides a properly normalized background
estimation. In Fig.~\ref{gkkb}b), there are 3169 selected $\gamma K^0_SK^0_S$
events and 413 background events.

\section{Analysis results}

\begin{figure}[htbp]
 \centerline{\epsfig{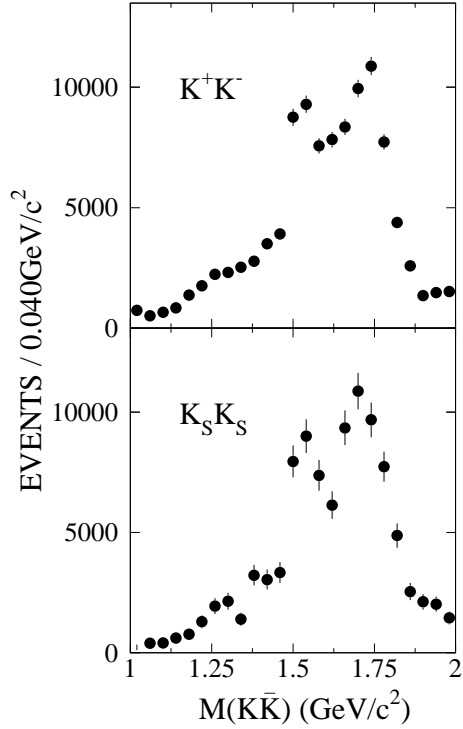}}
\caption{\label{fig:epsart1} The mass distributions of $K\bar K$ from
$J/\psi$ radiative decays, after acceptance and isospin corrections
for missing $\gamma K^0_L K^0_L$ and $\gamma K^0_S K^0_S$ with $K^0_S \to
\pi^0 \pi^0$ decays.}
  \label{corr-ggkk}
\end{figure}

We have carried out partial wave analyses using relativistic covariant
tensor amplitudes constructed from Lorentz-invariant combinations of
the 4-vectors and the photon polarization for $J/\psi$ initial states
with helicity $\pm 1$ \cite{GZJ}. Cross sections are summed over
photon polarizations.  The relative magnitudes and phases of the
amplitudes are determined by a maximum likelihood fit. The background
events obtained from Monte Carlo simulation or $\delta^2_{K_S}$ side
band are included into the data samples, but with the opposite sign of
log likelihood compared to data. These events cancel background within
the data samples. These analyses are confined to masses less than 2
GeV in order to ensure that a description containing only $0^{++}$ and
$2^{++}$ amplitudes be appropriate. The mass distributions of $K^+K^-$
and $K^0_SK^0_S$ after acceptance and isospin corrections for missing $\gamma
K^0_L K^0_L$ and $\gamma K^0_S K^0_S$ with $K^0_S \to \pi^0 \pi^0$
decays are shown in Fig. \ref{corr-ggkk}. The event topologies of the
$K^+K^-$ and $K^0_SK^0_S$ modes are different, so that acceptance and
background effects are rather different also; nevertheless, there is
good quantitative agreement between the two distributions.

\subsection{Bin-by-bin analysis}
\begin{figure}[ht]\includegraphics[height=140mm,width=120mm]{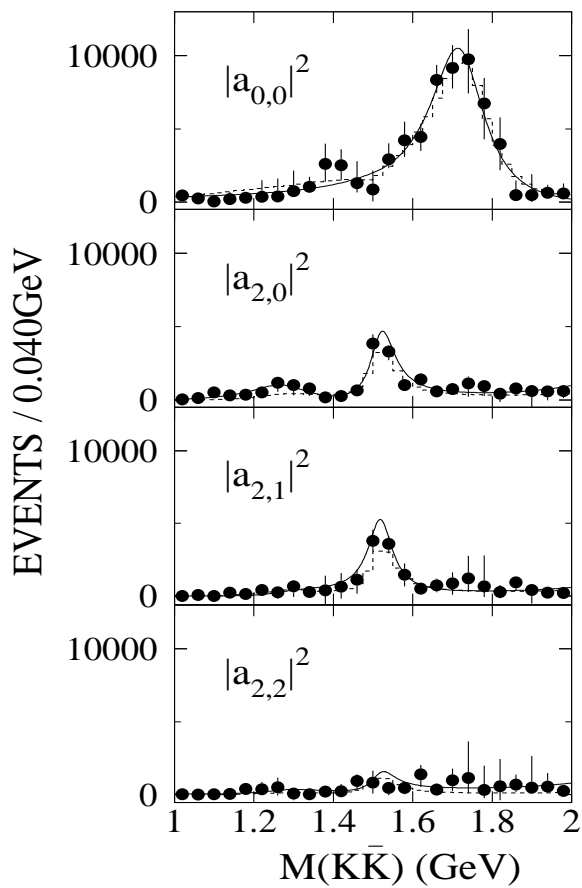}
\vspace*{-40pt}\caption{\label{fig:epsart2} The mass dependence of the
amplitude intensities for $\gamma K\bar K$ data. The solid curves
correspond to the coherent superposition of the Breit-Wigner
resonances fitted to the acceptance- and isospin-corrected data points
obtained from the bin-by-bin fit. The dashed line histograms are the
results of the global fit described in the text.}
  \label{heli}
\end{figure}

In the bin-by-bin analysis, the data in mass intervals 40 MeV wide
are fitted with four helicity amplitudes, one for $J^P=0^+$ and three for
$2^+$ amplitudes \cite{WMD}. The mass interval width is chosen as a compromise
between the desire for high statistics in each mass interval, and the need
for detailed information on the mass dependence of each measured
amplitude.  In each mass interval, the $\gamma K \bar K$ data sample
is analyzed in terms of the joint production and decay
angular distribution of the pseudoscalar meson system. The S- and
D-wave intensity distributions, $|a_{0,0}|^2$, $|a_{2,0}|^2$,
$|a_{2,1}|^2$ and $|a_{2,2}|^2$ for $\gamma K\bar K$ data resulting
from this bin-by-bin fit are shown as a function of mass in Fig. \ref{heli}.

The $K\bar K$ S-wave intensity dominates the 1.7 GeV region.
The solid curves in Fig. \ref{heli} correspond to fits of coherent
superpositions of individual Breit-Wigner resonances to the data
points of each intensity distribution.  The following channels are considered:
\begin {eqnarray*} J/\psi &\to& \gamma f'_2(1525) \\
       &\to& \gamma f_0(1710) \\
       &\to& \gamma f_2(1270) \\
       &\to& \gamma f_0(1500) \\
       &\to& \gamma + {\rm broad~} 0^{++} {\rm~and~} 2^{++} {\rm ~components}
\end {eqnarray*}
The first two are dominant. There is evidence for existence of the
$f_2(1270)$, and the $f_0(1500)$ is included here for consistency with
the global fit below.

For the spin 0 amplitude, two interfering resonances ($f_0(1500$,
$f_0(1710)$) and an interfering constant amplitude term, which is
used to describe the broad S- wave contribution, are
included.  The mass and width of the $f_0(1500)$ are fixed to the PDG
values; those of the $f_0(1710)$ are to be
determined.  The $f_0(1710)$ is well described by a Breit-Wigner of
mass and width M $= 1722\pm 17$ MeV, $\Gamma = 167^{+37}_{-29}$ MeV,
and the branching fraction for $J/\psi$ radiative decay to the
combined $K\bar K$ modes is ${\cal B}(J/\psi\to\gamma
f_0(1710)\to\gamma K\bar K) =(11.1^{+1.7}_{-1.2})\times 10^{-4}$.
The errors here are statistical errors.

For the spin 2 amplitudes, the $f'_2(1525)$ and $f_2(1270)$ are
included. There is also some $2^{++}$
structure above 2.0 GeV in $K\bar K$ mass, which could contribute to
the present fitted range, and thus the tail of a high mass $2^{++}$ state
is included in our fit.  We choose a resonance mass of 2250 MeV and
width of 350 MeV to represent the structure in the higher mass
region. The mass and width of the $f_2(1270)$ are
fixed at the values quoted in the PDG.  For the tensor resonance,
$f'_2(1525)$, its mass and width are fixed to the values M = 1519 MeV,
$\Gamma = 75$ MeV determined by the global fit which is
described below, and the total branching
fraction and ratios of  amplitude intensities are
determined to be ${\cal B}(J/\psi\to\gamma f'_2(1525)\to\gamma K\bar K) =
(4.02\pm 0.51)\times 10^{-4}$;
$x^2\equiv{\displaystyle{{|a_{2,1}|^2}/{|a_{2,0}|^2}}}=1.32\pm0.29$,
$y^2\equiv{\displaystyle{{|a_{2,2}|^2}/{|a_{2,0}|^2}}}=0.38\pm0.20$.
The intensity of the $f_2(1270)$ is poorly measured because of the
relatively low statistics and the weak coupling of this state to
$K\bar K$. The amount of spin 2 component in the 1.7 GeV mass
region is small, $\sim (16\pm9)\%$.  The errors shown above are
statistical and are obtained from the Breit-Wigner fit.

\subsection{Global fit analysis}

We now turn to the global fit to the $J/\psi\to \gamma K^+K^-$ and
$J/\psi\to\gamma K^0_SK^0_S$ data. Each sample is analyzed
independently, and the fit results shown below are for their averaged values.
This fit has the merit of constraining phase variations as a function of
mass to simple Breit-Wigner forms. It also performs the optimum
averaging of helicity amplitudes and their phases over resonances.
Partial waves are fitted to the data for the same components described in
the bin-by-bin fit.
The broad $0^{++}$ component improves the fit significantly; removing it
causes the log likelihood value to become worse by 221.
For the $f_2(1270)$ and $f_0(1500)$, we use PDG values of masses
and widths, but allow the amplitudes to vary in the fit.
For the $f_2'(1525)$, relative phases are consistent with zero within
experimental errors.
It is expected theoretically that relative phases should be
very small, on order of $\alpha \simeq 1/137$ for the electromagnetic
transitions
$J/\psi \to \gamma + 2^+$.
In view of the agreement with expectation, these relative phases
are set to zero in the final fit, so as to constrain
intensities further.

\begin{figure}
 \centerline{\epsfig{file=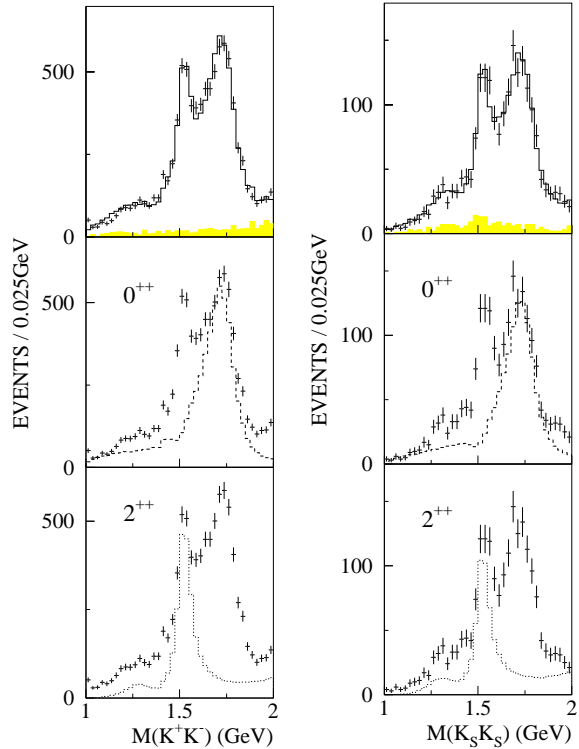,width=75mm}}
\caption{\label{fig:epsart3}
The $K\bar K$ invariant mass distributions from
$J/\psi \to \gamma K^+K^-$ and $J/\psi \to \gamma K^0_SK^0_S$.
The points are the data and the full histograms in the top panels show
the maximum likelihood fit. Histograms on subsequent panels show
the complete $0^+$ and $2^+$ contributions including all interferences.}
  \label{gkkfit}
\end{figure}

A free fit to $f_2'(1525)$ gives a fitted mass of $1519 \pm 2$ MeV and
a width of $75 \pm 4$ MeV. The fitted mass and width of the
$f_0(1710)$ are M $=1740\pm4$ MeV and $\Gamma = 166^{+5}_{-8}$ MeV,
respectively.  The fitted intensities are illustrated in Fig. \ref{gkkfit}.
For the $f_2'(1525)$, we find the ratios of helicity amplitudes $x^2 =
1.00\pm0.28$ and $y^2 = 0.44\pm0.08$.  In this fit, we allow some
$0^{+}$ contribution under the $f_2'(1525)$ peak, while previous
analyses by DM2 and Mark\,III [10, 11] ignored the small $0^+$
contributions.  The branching fractions of the $f'_2(1525)$ and the
$f_0(1710)$ determined by the global fit are ${\cal B}(J/\psi\to\gamma
f'_2(1525)\to\gamma K\bar K) = (3.42\pm0.15)\times 10^{-4}$ and ${\cal
B}(J/\psi\to\gamma f_0(1710)\to\gamma K\bar K) = (9.62\pm0.29)\times
10^{-4}$ respectively. The errors shown here are also statistical.
An alternative fit to $f_J(1710)$ with $J^P = 2^+$ 
is worse by 258 in log likelihood relative to $0^+$ for $\gamma K^+K^-$ data and by 67 for
$\gamma K^0_SK^0_S$.  Remembering that three helicity amplitudes are
fitted for spin 2 but only one for spin 0, the fit with $J^P = 0^+$ is
preferred by $> 10\sigma$ after considering the two data samples
together.

\begin{figure}
 \centerline{\epsfig{file=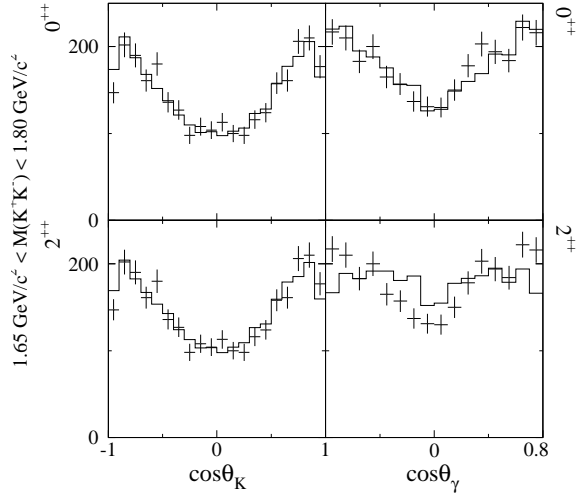,width=75mm}}
\caption{\label{fig:epsart4}
Projections in $\cos\theta_{K}$ and
$\cos\theta_{\gamma}$ for $0^{++}$ and $2^{++}$ assumptions. The points
are
the data ($J/\psi\to\gamma K^+K^-$ sample), and the histograms are the
global fit results.}
  \label{sepa}
\end{figure}

The separation between spin 0 and 2 is illustrated in Fig. \ref{sepa}, taking
the $J/\psi\to\gamma K^+K^-$ data as the example.
Let us denote the polar angle of the kaon in the $K\bar K$ rest frame
by $\theta_{K}$, and the polar angle of the photon in the
$J/\psi$ rest frame by $\theta _{\gamma }$.
The data are fitted simultaneously
including important correlations between $\theta_{K}$ and
$\theta _{\gamma }$. The left panels show resulting fits to
$\cos \theta_{K}$ for $J = 0$ and 2. There is no significant
difference between the two fits. The distributions should be flat
for $0^+$, but the interference with the tail of $f_2'(1525)$ has
a large effect. The right panels show the fits
to $\cos \theta _{\gamma }$; the optimum fit is visibly
better for $J = 0$ than for $ J = 2$. [If one fits {\it only} the
$\cos \theta _{\gamma}$ distribution, it is possible to fit equally
well with $J = 0$ or 2, but then the fit to
$\cos \theta_{K}$ gets much worse.]

If the $f_0(1500)$ is removed from the fit, the log
likelihood is worse by 1.65 (3.58) for $K^+ K^-$ ($K^0_S K^0_S$),
corresponding to about 1.3$\sigma$ (2.2$\sigma$).  If the $f_2(1270)$
is removed, the likelihood is worse by 57.5 (13.6) for $K^+ K^-$
($K^0_S K^0_S$), corresponding to $> 5 \sigma$ (3.8$\sigma$).

\section{Systematic error}
The systematic error for the global fit is estimated by adding or
removing small components used in the fit, replacing the $f_0(1500)$
with the $f_0(1429)$, $\Gamma = 169$ MeV, described in Ref.
\cite{WMD}, varying the mass and width of the large $f'_2(1525)$
within the PDG errors, varying the mass and width of $f_0(1710)$ based
on the difference between the $K^+K^-$ and $K^0_SK^0_S$ decay modes,
and varying the background component within reasonable limits in both
the global fit and bin-by-bin fit.  It also includes the uncertainty
in the number of $J/\psi$ events analyzed and the difference from two
 different choices of MDC 
wire resolution simulation.

The uncertainty about the
shape of broad $0^{++}$ background is included in the systematic error
also. An incoherent fit with this broad component and a fit with
alternative forms for the $s$-dependence using the parametrization of
Zou and Bugg \cite{ZOU} for the $f_0(400-1200)$ have been performed to
estimate the systematic error from this source.  This uncertainty
affects the results significantly, especially the branching fractions,
because of the interference between the broad structure and the other
components. Therefore, the error from this model-dependence for the
branching fraction measurements is separated from the statistical and
other systematic errors in our final results. The systematic errors
for the global fit are summarized in Table \ref{sys}. For the
mass and width, only the contributions from the model-dependence,
which are large compared to the other errors, are shown in the
table.

{\small
\begin{table}[htbp]
\begin{center}
\begin{tabular}{l|cccccccr}
\hline
&~M$_{f'_2(1525)}$~ &~$\Gamma_{f'_2(1525)}$~ &$x^2$&$y^2$&${\cal
  B}_{f'_2(1525)}$~&~M$_{f_0(1710)}$~
&~$\Gamma_{f_0(1710)}$~ &~${\cal
B}_{f_0(1710)} $~\\
\hline
remove $f_0(1500)$&&&~$^{+32}_{-0}$~~&~$^{+20}_{-0}$~~&$\pm0$&&&$^{+10}_{-0}$\\
use $f_0(1429)$&&&$^{+0}_{-15}$&$^{+0}_{-9}$&$^{+0}_{-5}$&&&$^{+3}_{-0}$\\
remove $f_2(1270)$&&&$^{+42}_{-0}$&$^{+0}_{-55}$&$^{+6}_{-0}$&&&$^{+0}_{-1}$\\
use the $\sigma$&$+0.66$&$+20$&$^{+17}_{-9}$&$^{+0}_{-14}$&$^{+33}_{-0}$&$-1.44$&$+9$&$^{+29}_{-0}$\\
incoherent $0^{++}$&$+0.99$&---&$^{+6}_{-0}$&$^{+0}_{-64}$&$^{+45}_{-0}$&---&$+3$&$^{+28}_{-0}$\\
M, $\Gamma$ of $f'_2(1525)$&&&$^{+49}_{-15}$&$^{+0}_{-34}$&$^{+11}_{-8}$&&&$^{+4}_{-5}$\\
M, $\Gamma$ of $f_0(1710)$&&&$^{+51}_{-17}$&$^{+11}_{-36}$&$\pm3$&&&$^{+1}_{-0}$\\
M, $\Gamma$ of high $2^{++}$&&&$^{+46}_{-14}$&$^{+0}_{-59}$&$^{+1}_{-4}$&&&$^{+6}_{-0}$\\
background&&&$^{+46}_{-17}$&$^{+0}_{-55}$&$^{+0}_{-3}$&&&$^{+9}_{-10}$\\
$\delta_{N_{J/\psi}}$&&&---&---&$\pm4.7$&&&$\pm4.7$\\
wire resolution &&&---&---&$\pm15$&&&$\pm15$\\
\hline
\end{tabular}
\caption{Estimation of systematic error (\%) in the global fit. ${\cal
  B}_{f'_2(1525)}$ and ${\cal
  B}_{f_0(1710)}$ are the branching fractions for $f'_2(1525)$ and $f_0(1710)$ respectively.}
\label{sys}
\end{center}
\end{table}
}

\section{Results and Discussion}
The results of the bin-by-bin and global fits are summarized in
Tables \ref{res1} and \ref{res2} respectively. For the bin-by-bin fit, the errors are
statistical ones only, and for the global fit, the first error listed is the statistical
error, the second error is the systematic error, and the third one for the branching fractions is
for the model-dependence of the broad components.

\begin{table}[htbp]
\begin{tabular}{l|cr}\hline
&$f'_2(1525)$&$f_0(1710)$\\
\hline
M (MeV)& ~~1519 (fixed)~~ &~~$1722\pm17$~~\\
\hline
$\Gamma$
(MeV)&75 (fixed)&$167{^{+37}_{-29}}$\\
\hline
${\cal B}(J/\psi\to\gamma X,$&&\\
$X\to K \bar K)(\times 10^{-4}) $&$4.02\pm0.51$&
$11.1^{+1.7}_{-1.2}$\\
\hline
$x^2={\displaystyle{{|a_{2,1}|^2}/{|a_{2,0}|^2}}}$&$1.32\pm0.29$&-----\\
\hline
$y^2={\displaystyle{{|a_{2,2}|^2}/{|a_{2,0}|^2}}}$&$0.38\pm0.20$&-----\\ \hline
\end{tabular}
\caption{\label{tab:table2}
Measurements of the $f'_2(1525)$ and $f_0(1710)$ for the bin-by-bin
fit. Errors shown are statistical only.}
\label{res1}
\end{table}

\begin{table}[htbp]
\begin{tabular}{l|cr}\hline
&$f'_2(1525)$&$f_0(1710)$\\
\hline
M (MeV)&$1519\pm2{^{+15}_{-5}}$&$1740\pm4{^{+10}_{-25}}$\\
\hline
$\Gamma$ (MeV)&$75\pm4{^{+15}_{-5}}$&$166{^{+5}_{-8}}{^{+15}_{-10}}$\\
\hline
${\cal B}(J/\psi\to\gamma X,$&&\\
$X\to K \bar K)(\times10^{-4}) $&~~$3.42\pm0.15{^{+0.69}_{-0.65}}{^{+1.55}_{-0.00}}$~~&
~~$9.62\pm0.29{^{+2.11}_{-1.86}}{^{+2.81}_{-0.00} }$~~\\
\hline
amp. ratios $x^2$&$1.00\pm0.28{^{+1.06}_{-0.36}}$&-----\\
\hline
~~~~\,~~~~\,~~~~~~~$y^2$&$0.44\pm0.08{^{+0.10}_{-0.56}}$&-----\\\hline
\end{tabular}
\caption{\label{tab:table1}
Measurements of the $f'_2(1525)$ and $f_0(1710)$ for the global
fit. The first error is statistical, the second is systematic, and the
third is that corresponding to model-dependence of the broad components.}
\label{res2}
\end{table}

The two fit methods, bin-by-bin  and global, are based on different
analysis concepts. In the bin-by-bin fit, the S- and D-wave
intensities are fairly well determined and nearly model independent.
The only model dependence in the bin-by-bin fit is the assumption that
only S- and D-waves need be considered; this is reasonable, since one
would not expect significant $4^{++}$ amplitudes below 2 GeV. However,
due to limited statistics for each bin and the limited solid angle
coverage of the detector, the relative phases of partial waves cannot
be well determined.  This causes larger uncertainties when extracting
the mass and width of resonances by fitting only the partial wave
intensities without the constraints of the relative phases between them.
In the global fit, the phase variations as a function of mass are
constrained to simple Breit-Wigner (BW) forms .  The stability of the
minimum optimizing procedure and statistical errors are better than
those of the bin-by-bin fit.  However, if some non-BW resonance is
assumed to be a BW-form amplitude, this will give a model-dependent
biased result. The model independent bin-by-bin result for the partial
wave intensities can provide guidance for choosing components for
the global fit.  The final full amplitudes from the global fit
definitely give a better fit to the whole set of data than the
amplitudes obtained from fitting the partial wave intensities without
constraints of relative phases between them.
 
Fortunately from Tables \ref{res1} and \ref{res2} and the comparison
shown in Fig. \ref{heli}, we see that the results obtained from the
bin-by-bin fit and 
the global fit for the $f'_2(1525)$ and $f_0(1710)$
agree with each other well within the errors.  The ratios of the
helicity amplitudes of the $f'_2(1525)$ from the present analysis are
in reasonable agreement with Krammer's predictions
\cite{krammer}. These ratios provide useful information for testing
models of the resonance production and decay mechanisms. Most
importantly, the analysis demonstrates that the mass region around 1.7
GeV is predominantly $0^{++}$ from the $f_0(1710)$ \cite{2pp}; this
conclusion is consistent with that of references [3-5].

\section{Summary}

    In summary, the partial wave analyses of $J/\psi\to\gamma K^+K^-$
and $J/\psi\to\gamma K^0_SK^0_S$ using 58M $J/\psi$ events of BES\,II
show strong production of the $f'_2(1525)$ and the S-wave resonance
$f_0(1710)$. This confirms earlier conclusions that the spin-parity of
the $f_0(1710)$ is $J^P = 0^+$.  The $f_0(1710)$ peaks at a mass of
$1740\pm 4^{+10}_{-25}$ MeV with a width of
$166{^{+5}_{-8}}{^{+15}_{-10}}$ MeV.  For the $f'_2(1525)$, the
helicity amplitude ratios are determined to be
$1.00\pm0.28^{+1.06}_{-0.36}$ and $0.44\pm{0.08}^{+0.10}_{-0.56}$,
respectively.  They are consistent with theoretical predictions.

\section{Acknowledgments}

\vspace{0.4cm}

The BES collaboration acknowledges the strong efforts of the BEPC
staff and the helpful assistance we received from the members of the
IHEP computing center.  We also wish to thank William Dunwoodie and
Walter Toki for useful discussions and suggestions. This work is
supported in part by the National Natural Science Foundation of China
under contracts Nos. 19991480,10225524,10225525, the Chinese Academy
of Sciences under contract No. KJ 95T-03, the 100 Talents Program of
CAS under Contract Nos. U-11, U-24, U-25, and the Knowledge Innovation
Project of CAS under Contract Nos. U-602, U-34(IHEP); by the National
Natural Science Foundation of China under Contract No.10175060(USTC);
and by the Department of Energy under Contract No.
DE-FG03-94ER40833 (U Hawaii).  We wish to acknowledge financial
support from the Royal Society for collaboration between the BES group
and Queen Mary College, London.


\begin{thebibliography}{99}
\bibitem{QCDL} G. Bali, K. Schilling, A. Hulsebos, A. Irving, C.
Michael, and P. Stephenson, Phys. Rev. {\bf B309} (1993) 378;
 C. Michael, Proceedings of Hadron 97, AIP
Conf. Series {\bf 432} (1997) 657;
 W. Lee, and D. Weingarten, hep-lat/9805029;
 C. Morningstar, and M. Peardon, Phys. Rev. {\bf D60}
(1999) 034509.
\bibitem {PDG} K. Hagiwara $et~al.$ (Particle Data Group),
Phys. Rev. {\bf D66} (2002) 010001.
\bibitem {WMD} W. Dunwoodie, {\it Hadron Spectroscopy}, AIP Conf. Series
             {\bf 432} (1997) 753.
\bibitem {WA76} B. French $et~al.$, Phys. Lett. {\bf B214} (1999) 213.
\bibitem {WA102} D. Barberis $et~al.$, Phys. Lett. {\bf B453} (1999) 305 and
316;
Phys. Lett. {\bf B462} (1999) 462.
\bibitem{BESII} J.Z. Bai $et~al.$ (BES Collaboration),
Nucl. Instr. Meth. {\bf A458} (2001) 627.
\bibitem{THimel} T. Himel $et~al.,$ Phys. Lett. {\bf
45} (1980) 1146.
\bibitem {JINS} J.Z. Bai $et~al.$ (BES Collaboration), Phys. Rev. Lett. {\bf 76} (1996) 3502.
\bibitem {GZJ} B.S. Zou and D.V. Bugg, Eur. Phys. J. {\bf A16} (2003) 537.
\bibitem {DM2} J.E. Augustin $et~al.$, Zeit. Phys. {\bf C36} (1987) 369.
\bibitem {MK3} R.M. Baltrusaitis $et~al.$, Phys. Rev. {\bf D35} (1987) 2077.
\bibitem {ZOU} B.S. Zou and D.V. Bugg, Phys. Rev. {\bf D48} (1993)
3948.
\bibitem {krammer} M. Kramer, Phys. Lett. {\bf B74} (1978) 361.
\bibitem{2pp} The amount of the possible $2^{++}$ component is even
smaller in the 1.7 GeV mass region in the global fit, less than
$10\%$.  

\end{thebibliography}

\end{document}